\documentclass{sig-alternate-10pt}
\usepackage[margin=1in]{geometry}
\usepackage{url}

\usepackage{breakurl}
\usepackage[breaklinks]{hyperref}
\usepackage{xspace}
\usepackage{cite}
\usepackage{multirow}
\usepackage{balance}
\usepackage{enumitem}

\usepackage{xcolor}
\usepackage{colortbl}
\definecolor{c_blue}{HTML}{0173B2}

\newcommand{\dq}[1]{\textsf{#1}\xspace} 
\newcommand{\mypara}[1]{\smallskip\noindent\textbf{#1.}}
\usepackage[numbers]{natbib}

\usepackage{challenges}
\newcommand{\addsingletablerow}[5]{%
    \begin{center}
    \begin{tabular}{c|c|c|c|c}
    \multicolumn{5}{c}{}\\[-1.em]
    Data & Source & System & Task & Human \\ \hline
    #1 & #2 & #3 & #4 & #5 \\ \multicolumn{5}{c}{}\\[-1.em]
    \end{tabular}
    \end{center}
}

\newcommand{\revision}[1]{#1}

\usepackage{titlesec}
\titlespacing*{\section}{0pt}{1.0\baselineskip}{0.25\baselineskip}
\begin{document}
\emergencystretch 1em
\pagestyle{empty}

\title{The Five Facets of Data Quality Assessment}

\author{
 Sedir Mohammed\textsuperscript{1}, Lisa Ehrlinger\textsuperscript{1}, Hazar Harmouch\textsuperscript{2}, Felix Naumann\textsuperscript{1}, Divesh Srivastava\textsuperscript{3}\\ \affaddr{\textsuperscript{1}Hasso Plattner Institute, University of Potsdam, Germany}\\ \affaddr{\textsuperscript{2}University of Amsterdam, Netherlands}\\ \affaddr{\textsuperscript{3}AT\&T Chief Data Office, USA}\\ \email{sedir.mohammed@hpi.de, lisa.ehrlinger@hpi.de}\\ \email{h.harmouch@uva.nl, felix.naumann@hpi.de, divesh@research.att.com}
}

\maketitle

\begin{abstract}
Data-oriented applications, their users, and even the law require data of high quality.
Research has divided the rather vague notion of data quality into various dimensions, such as \dq{accuracy}, \dq{consistency}, and \dq{reputation}. 
To achieve the goal of high data quality, many tools and techniques exist to clean and otherwise improve data.
Yet, systematic research on actually assessing data quality in its dimensions is largely absent, and with it, the ability to gauge the success of any data cleaning effort.

We propose five facets as ingredients to assess data quality: \emph{data}, \emph{source}, \emph{system}, \emph{task}, and \emph{human}.
Tapping each facet for data quality assessment poses its own challenges.
We show how overcoming these challenges helps data quality assessment for those data quality dimensions mentioned in Europe's AI~Act.
Our work concludes with a proposal for a comprehensive data quality assessment framework.




\end{abstract}

\section{The Many Dimensions of Data Quality}
\label{sec:intro}

\textit{Data quality}~(DQ) has been an important research topic for the past decades~\cite{batini_data_2016,DBLP:conf/iq/NaumannR00,DBLP:journals/jmis/WangS96}, reflecting its critical role in all fields where data are used to gain insights and make decisions. 
A manifold of DQ dimensions exists that regard data and their properties from various perspectives and contribute to understanding and characterizing the complex nature of data~\cite{batini_data_2016,DBLP:journals/jmis/WangS96}.
\begin{figure}[t]
    \centering
    \includegraphics[width=0.83\linewidth]{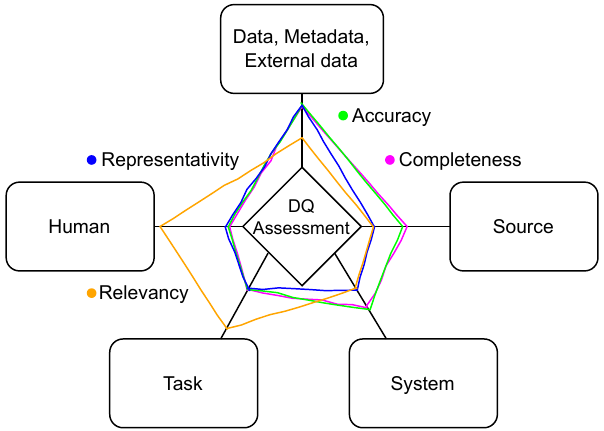}
    \caption{The five \emph{facets} of DQ assessment and exemplary characteristics for DQ dimensions.}
    \label{fig:facets_overview}
\end{figure}

\mypara{The high demand for DQ} Especially in the fast-moving landscape of \textit{artificial intelligence}~(AI), where data plays a pivotal role, the significance of DQ is dramatically increasing, so much so that literature calls this trend a paradigm shift from a model-centric view to a data-centric one~\cite{DBLP:journals/corr/abs-2303-10158}.
Data-centric AI emphasizes the data and their impact on the underlying model~\cite{neutatz2021cleaning, neutatz2022data, DBLP:journals/vldb/WhangRSL23}.
Literature showed that DQ, with its various dimensions, significantly influences prediction accuracy~\cite{neutatz2022data, budach2022effects, DBLP:conf/icde/ForoniLV21,DBLP:conf/icde/LiRBZCZ21}.
Domain-specific particulars provide a context that imposes specific requirements on  DQ assessment, such as the \textit{Health Insurance Portability and Accountability Act}~(HIPAA), 
which focuses on privacy but promotes DQ dimensions, such as \dq{accuracy} and \dq{completeness} for ensuring trust~\cite{noauthor_hipaa_2023}.

Such requirements have also become part of regulation, as in the \textit{General Data Protection Regulation}~(GDPR)~\cite{General_Data_Protection_Regulation} and the EU \textit{AI Act}~\cite{european_parliament_artifical_2024}. 
For instance, the AI Act mentions in \textit{Article 10} the DQ dimensions \dq{representativity}, \dq{accuracy}~(free of errors), \dq{completeness} and \dq{relevancy}~\cite{european_parliament_artifical_2024}.
Similar initiatives to regulate DQ and AI are also being made by the United States~\cite{house_executive_2023} and China~\cite{roberts_chinese_2021}, which underlines the international interest in the topic of DQ. 

Examining DQ is by no means just an academic problem~\cite{DBLP:journals/csur/BatiniCFM09}.
Industry is also concerned about the impact of DQ on business~\cite{sadiq_handbook_2013}.
Companies have shifted from internal ``data gazing''~\cite{maydanchik_data_2007} to hiring auditing firms for quality assurance.
Literature shows that poor DQ has an enormous economic impact on organizations, either through loss of revenues or through additional internal costs~\cite{nagle_assessing_2020,redman2001data}.

In addition to recognizing the relevance of DQ and understanding it in terms of the various dimensions, the goal is to improve DQ by cleaning the data.
Yet, quality cannot be improved if it cannot be measured~\cite{DBLP:journals/jasis/StviliaGTS07}:
we need concrete \emph{assessment methods} to evaluate DQ in individual dimensions.
\citet{DBLP:journals/csur/BatiniCFM09} define DQ assessment as the measurement of DQ and the comparison with reference values for diagnosing it.
As such, apart from the pure measurement of DQ, assessment includes classifying whether the measured quality is \emph{sufficient} (or ``fit'') for the underlying task. 
\emph{Measuring} vs.\ \emph{judging} whether the measured DQ suffices for a task at hand are challenges of rather different natures. 
\mypara{\revision{Vision statement}}
\revision{Given a dataset, a use case (task specification), a set of DQ dimensions, and their formal definitions, our goal is to develop effective and efficient assessment procedures for each DQ dimension. 
These procedures should compute values that accurately align with the formal definitions.}

\mypara{\revision{Mission statement}}
\revision{
To achieve the vision, 
we want to identify facets upon which  assessment procedures across DQ dimensions depend. 
These facets enable individual dimensions to benefit from solutions to shared assessment challenges and streamlined implementation of assessment procedures.}

\mypara{Contribution}
This paper proposes a new perspective on DQ research: through the lens of so-called \emph{facets}.
We discuss five \emph{facets} of DQ assessment as potential sources for DQ information.
Each \emph{facet} presents its own set of challenges and opportunities.
To overcome the challenges and capitalize on the opportunities, we identify a wide range of technologies that require cross-community expertise.
We envision the thorough implementation of these technologies by different research communities.
The ultimate goal is the integration of these technologies into a robust framework.
We advocate the development of a \textit{DQ assessment framework} for correctly and efficiently measuring all DQ dimensions.
The framework enables (1)~the integration of deeper data profiling methods~\cite{Abedjan15}, (2)~compliance with given regulations, (3)~enhancement of data cleaning, as well as (4)~\textit{judging} whether DQ meets user expectations.
While this paper focuses on structured data, we believe it can also be extended to semi-structured or unstructured data.







\section{Data Quality Assessment by Facets}
\label{sec:facets}

Data quality~assessment in its variety of dimensions~\cite{batini_data_2006, DBLP:conf/iq/NaumannR00} poses many definitional, computational, and organizational challenges.
We propose five \emph{facets}~(see Figure~\ref{fig:facets_overview}) that serve as foundation for DQ assessment: (i)~the \hyperref[sec:data_facet]{\emph{data} itself, including metadata and external data}; (ii)~the \hyperref[sec:source_facet]{\emph{source}} of the data; (iii)~the \hyperref[sec:system_facet]{\emph{system}} to store, handle, and access the data; (iv)~the \hyperref[sec:task_facet]{\emph{task}} to be performed on the data; and (v)~the \hyperref[sec:human_facet]{\emph{humans}} who interact with the data.
\revision{These five facets are inspired by the stages of a typical data life cycle~\cite{Stodden20}: all relevant components of each stage can be mapped to one or more facets.}

Each \emph{facet} poses its own challenges and opportunities for future research.
\revision{We hypothesize that addressing these challenges per \emph{facet} adresses problems that arise from more than one DQ dimension.}
We propose \emph{facets} as an additional layer to structure DQ research, allowing all dimensions involved in the assessment of a specific \emph{facet} to benefit simultaneously from solving these challenges.

In the following, we define and discuss each of the five \emph{facets} and their key challenges. We list exemplary DQ dimensions (see \cite{mohammed2024data} for definitions) that specifically benefit from resolving these challenges. 


\subsection{The Data Facet}\label{sec:data_facet}
Raw data values are intended to represent real-world concepts and entities.
The data facet includes the data semantics and their digital representation.
It also includes metadata, such as schema information and other documentation, and any assessment-relevant external knowledge (as data), like a knowledge base \revision{(e.g., DBpedia~\cite{DBpedia15})} to validate data.
The data facet encompasses all challenges related to the data being assessed, its metadata, and external data.

As data occur in different \emph{granularity}~(e.g., values, records, columns), DQ assessment must identify the necessary level of detail and devise quality-metric aggregation methods to cross levels of granularity.
Also, \textit{metadata}, such as schema and data types, should be available and of high quality itself.
When external knowledge is needed, challenges arise in discovering, matching, and assessing the quality of \textit{reference data}.
If data is encrypted, it cannot be assessed directly, so DQ assessment must handle \textit{encrypted data} and, in case of distribution, also work in a \textit{federated setup}.

In the following, we highlight two well-known DQ dimensions (mentioned in the AI Act) where the data facet is involved in the assessment and which specifically benefit from solving its challenges.
\challengebox{Example DQ Dimensions}{
\item \underline{\dq{Accuracy}:} Typical metrics to assess \dq{accuracy} require \textit{reference data} to determine how closely the data matches the reality. 
\item \underline{\dq{Completeness}:} Placeholders represent missing values, using either obvious placeholders like ``NaN'' or less obvious placeholders. The assessment needs \textit{metadata} that contains information about the placeholder representation.
}

\subsection{The Source Facet}\label{sec:source_facet}
The source of data represents a logical perspective. 
This \emph{facet} encompasses evaluating the data generation and collection processes, as well as assessing the source's integrity and organizational compliance. 
The main aspect of the source facet is data \emph{provenance}, which includes information on the origins, providers, and other organizations involved in creating and transforming the data~\cite{DBLP:journals/vldb/HerschelDL17}.

One key challenge is ensuring \textit{data lineage} traceability, including the data origin and its transformations~\cite{DBLP:conf/btw/GlavicD07}.
Additionally, a \textit{process-oriented view} is crucial, which includes evaluating the transformation process and the credibility of annotating agents in the DQ assessment.
It is also important to consider the \textit{time range} for assessing reliability over time; longer histories provide a more comprehensive view, while shorter intervals highlight recent changes. 
\challengebox{Example DQ Dimensions}{
\item \underline{\dq{Reputation}:} The assessment requires evaluating a data source's credibility and reliability, and thus, considering historical reliability with \textit{data lineage}.
\item \underline{\dq{Believability}:} The key challenge is to verify the data origin (\textit{time range}), source transformations (\textit{data lineage}), and involved entities (\textit{process}).
}

\subsection{The System Facet}\label{sec:system_facet}
The system facet pertains to a physical perspective, including the infrastructure and technology for storing, handling, and accessing the data.
It also covers the system's technical compliance with legal and regulatory requirements, ensuring adherence to necessary data management standards.

The system facet raises challenges, such as \textit{clarity} or \textit{auditability}.
The \textit{clarity} includes documenting the system's architecture, data processing capabilities, interoperability with other systems, security features, and user interface aspects.
\textit{Auditability} is crucial for verifying compliance with regulations, such as data deletion and security standards.
\challengebox{Example DQ Dimensions}{
\item \underline{\dq{Recoverability}:} Assessing the ability to restore a prior state of the data requires knowledge about the file system, backup procedures (\textit{clarity}) and long-term storage regulations (\textit{auditability}).
\item \underline{\dq{Portability}:} The key challenge is to understand the storage system, including file formats (\textit{clarity}) and interoperability standards (\textit{auditability}).
}

\subsection{The Task Facet}\label{sec:task_facet}
The task facet pertains to the specific use case and the context in which the data is employed.
Thus, it inherently aligns with the ``fitness for use'' definition of DQ~\cite{batini_data_2016,DBLP:journals/jmis/WangS96}.
The task influences which parts of the data~(e.g., columns, tuples) are considered and how well they represent the real-world.

The task facet poses challenges regarding the \textit{relevance} of the data, including the identification of relevant attributes and tuples. 
Also, the \textit{risk} of the task, according to the AI Act, which defines minimal-, limited-, high- and unacceptable-risk AI systems, can determine the way DQ is assessed~\cite{ai_act_risk_levels}.
Higher risk categories require more stringent DQ assessment methods, including strict validation processes and documentation, to ensure compliance.
\challengebox{Example DQ Dimensions}{
\item \underline{\dq{Timeliness}:} The key challenge is defining an acceptable timeframe for tasks and to classify how long data are considered up-to-date or \textit{relevant}.
\item \underline{\dq{Relevancy}:} The assessment involves balancing the need for complete information (\textit{relevance}) against the risk of including unnecessary data that can violate legal requirements (\textit{risk}).
}
    
\subsection{The Human Facet}\label{sec:human_facet}
The human facet \revision{introduces a subjective view, while including} the diverse groups that interact with the data, perform the task, and interpret the results.
It aligns DQ with the specific needs and contexts in which users operate.
Some DQ dimensions~(e.g., \dq{relevancy}, \dq{believability}), require user surveys to assess experiences and challenges in handling the data.
\revision{This subjective perspective makes it challenging to fully automate the assessment.}
The human facet poses challenges such as 
the need to \emph{design surveys} that capture a range of expertise levels, or also the consideration of the \textit{intent} of different user groups and their perspectives (e.g., developers, customers).
\challengebox{Example DQ Dimensions}{
\item \underline{\dq{Ease of manipulation}:} Since manipulability can impact accessibility positively and data integrity negatively, the assessment must consider the users \textit{intent} of manipulation.
\item \underline{\dq{Relevancy}:} Determining relevant data varies by user perspective (\textit{intent}). The evolving nature of \dq{relevancy} with changing user needs, market trends, and legal standards complicates maintaining up-to-date assessments (\textit{survey design}).
}
\section{Facet Application}
\label{sec:facets_application}

In the previous section, we listed example DQ dimensions per \emph{facet}, for which the considered \emph{facet} is involved in the assessment.
Of course, the participation of the \emph{facets} in assessing a DQ dimension occurs to varying degrees.
We use a three-level system~(``++'', ``+'', ``-'') to indicate a \emph{facets}' participation: ``++'' for strong involvement, ``+'' for medium, and ``-'' for low to no involvement.
We determined the involvement of the \emph{facets} through several discussion rounds among all authors until we reached a consensus.
When determining the involvement of \emph{facets}, we deliberately voted in favor of \revision{an objective and automatic assessment and thus tried to minimize the involvement of the human facet}.
\revision{Although DQ is often defined as ``fitness for use''~\cite{DBLP:journals/jmis/WangS96} the task facet is not necessarily included in the assessment.}

In the following, we 
discuss the \emph{facet} involvement and  implications with respect to specific technologies for each DQ dimension from the AI Act: \dq{accuracy}~(free of errors), \dq{representativity}, \dq{completeness}, and \dq{relevancy}~\cite{european_parliament_artifical_2024}~(see Figure~\ref{fig:facets_overview}).
\revision{Additionally, we include a discussion on \dq{accuracy} and \dq{relevancy} as examples to illustrate why certain facets are not involved in the assessment.}

\subsection{DQ Dimension: Accuracy}\label{sec:accuracy}
\paragraph{Definition} 
\dq{Accuracy} describes the correspondence between a phenomenon in the world and its description as data~\cite{batini_data_2016}.
\addsingletablerow{++}{+}{+}{-}{-}
The data facet is the primary contributor to the assessment of \dq{accuracy}.
Further aspects from the source facet (e.g., data provenance) and the system facet (e.g., storage technologies) are also relevant.
\revision{Conversely, the task and human facets are less relevant: \dq{accuracy} can be measured on a purely objective level, considering factual correctness and alignment with truth.}

Literature established several metrics to assess \dq{accuracy}~\cite{Haegemans_2016, DBLP:journals/csur/BatiniCFM09}.
Most metrics require reference data, which corresponds to the data facet.
To address this challenge, the reference data must be defined (e.g., its level of detail) and collected.
Open data platforms, such as Kaggle~\cite{noauthor_kaggle_nodate} 
or general knowledge bases (e.g., Wikidata~\cite{noauthor_wikipedia_nodate}, DBpedia~\cite{DBpedia15}), 
are well suited to collect a variety of data.
To make use of such external data, they must be matched with the data using \emph{schema matching} approaches~\cite{DBLP:journals/vldb/RahmB01, DBLP:journals/csur/BatiniL86, doan2012principles,herzog_data_2007}, which must handle different formats to process reference data from different sources~\cite{DBLP:conf/vldb/MiloZ98}.
This is particularly challenging with data that include \emph{natural language}, demanding methods for semantic and syntactic processing, potentially using \emph{large language models}~\cite{DBLP:journals/pvldb/FernandezEFKT23}.

In cases where access to such data platforms is too expensive or where no relevant data of sufficient quality could be found, \emph{semantic web technologies} combined with \emph{information retrieval approaches} would allow gathering data from the web, as external data for assessment~\cite{10.1007/3-540-48005-6_21, DBLP:conf/cikm/ShahFJ02, grossman_information_2004}.


In terms of the source facet, error detection and cleaning methods, such as NADEEF~\cite{DBLP:conf/sigmod/DallachiesaEEEIOT13} or HoloClean~\cite{DBLP:journals/corr/RekatsinasCIR17}, can be used to identify and correct data errors.
The transformations applied must be clearly documented in the metadata (see Section~\ref{sec:completeness}).

The system in which the data is stored might be responsible for erroneous values caused by system failures, such as crashes or bugs.
Thus, the system can lose information when saving new values, such as decimal points. 
Consequently, system robustness, data replication, and recovery processes must be included in the metadata.
These aspects require a cataloging system to format the metadata in a machine-readable format (see also Section~\ref{sec:completeness}).

The system in which the data and metadata are located must ensure that access to them aligns with the relevant privacy provisions.
If the data owner grants consent, where the consent information can also be part of the cataloging system, a partial decryption can be performed.
Alternatively, encryption schemes such as \emph{homomorphic encryption} can be used to assess and process the data/metadata while they are encrypted~\cite{DBLP:journals/csur/AcarAUC18}.
Compliance with privacy provisions is independent of the assessment of specific DQ dimensions.

\subsection{DQ Dimension: Representativity}\label{sec:representativity}
\paragraph{Definition}
\dq{Representativity} aims to ensure that the characteristics of the reference data are present in the considered data~\cite{10bd9b32-8b48-36d1-a9cc-51fbb4bf7d53, DBLP:journals/corr/abs-2203-04706}.
\addsingletablerow{++}{-}{-}{-}{-}
The data facet is the main contributor to the assessment of \dq{representativity}.

Similar to \dq{accuracy}, metrics to assess \dq{representativity} require information on the reference data~\cite{DBLP:journals/entropy/BudkaGM11, DBLP:journals/corr/abs-2203-04706}.
Thus, the reference data must first be defined to establish a baseline for comparison in the assessment.
In contrast to \dq{accuracy}, assessing \dq{representativity} does not require the complete reference data -- summary statistics, respectively, data distributions of the attributes, are often sufficient.
Depending on the data source, \emph{metadata} may already contain information about summary statistics and distributions.
This metadata must be in  a \emph{structured format} (e.g., JSON or RDF) to enable automated access and further processing.
Beyond uniform formatting, information must follow a \emph{uniform schema} and \emph{vocabulary} across data sources to ensure interoperability.
The use of an \emph{ontology} (e.g., Croissant~\cite{akhtar2024croissant} or DSD~\cite{dsd_2023}) would ensure a standardized schema and vocabulary, improving interoperability.

Still, the data must be matched with the given data, even if it is in an aggregated format.
But, \emph{data matching} with less data is an easier task because there are fewer records and attributes to compare, reducing computational complexity and processing time.
This simplifies schema matching, data cleaning, and handling diverse formats, leading to fewer errors and more straightforward and accurate matching criteria.
Nevertheless, if the external data sources do not provide this information, the technologies the assessment requires to obtain and match the reference data overlap with the technologies mentioned in the context of \dq{accuracy}.


\subsection{DQ Dimension: Completeness} \label{sec:completeness}
\paragraph{Definition}
\dq{Completeness} refers to the extent to which data, including entities and attributes, are present according to the data schema~\cite{10.1145/505248.506010}.
\addsingletablerow{++}{+}{+}{-}{-}
\revision{When focusing on entry-level \dq{completeness}, }the data facet is primarily involved in the assessment;
the source and system facets partially.

Since \dq{completeness} represents the presence of the data, its assessment requires the measurement of missing values.
While \texttt{null} or conventional placeholders like ``NaN'' for missing values are easily identified, more research is required to also identify so-called ``hidden missing values'' like ``-99'', ``EMPTY'', or default values~\cite{FAHES2018, Bechny_2021}.
Identifying these hidden missing values can either be done through prior knowledge~(in terms of metadata and sophisticated \emph{Data Catalogs}~\cite{DBLP:conf/dexaw/EhrlingerSMKW21} or, particularly suited for the ML context, with \emph{Data Cards}~\cite{DataCards22}) or alternatively learned with ML models taking into account the context.
Placeholders can differ for each data source or be domain-specific, which is why strict documentation is important.
In addition, transformations on missing values, like deleted records or applied imputation strategies, must also be part of the metadata.

Similar to \dq{accuracy}, the system in which the data is located might cause missing values, e.g., due to hardware failure.
In the context of \dq{completeness}, the system can lose data or fail to store new values, again necessitating metadata for recovery processes. 

\subsection{DQ Dimension: Relevancy}\label{sec:relevancy}
\paragraph{Definition}
\revision{\dq{Relevancy} describes the extent to which data are applicable and helpful for a given task}~\cite{DBLP:journals/jmis/WangS96}.
\addsingletablerow{+}{-}{-}{++}{++}
While the task and the human facet mainly support the assessment of \dq{relevancy}, the data facet is also involved.
\revision{Conversely, the source and system facets are less relevant, as \dq{relevancy} is solely determined by the data's usefulness for fulfilling a specific task, regardless of how or where it was created or stored.}

To assess \dq{relevancy}, stakeholders must \emph{define} the given task, requiring domain experts to incorporate best practices and to understand the task's intricacies. 
Given the task, stakeholders and experts have to assess the \dq{relevancy} of individual attributes and tuples.
Alternatively, \emph{statistical methods} can assess relevancy, e.g., Shapley or LIME calculate the feature importance to determine each feature's contribution to an ML model's prediction~\cite{shapley1953value, DBLP:conf/icml/SundararajanN20, DBLP:conf/nips/SlackHSL21}.
As feature importance is computationally complex, manual assessment might still be needed.

This manual assessment can be supported with \emph{data profiling}~\cite{DBLP:journals/sigmod/Naumann13} methods, comprising several tasks, such as, the automatic identification of distributions, functional dependencies, or data types. 
Based on the gathered information, experts can define domain- and task-specific criteria to assess the relevance of individual attributes and tuples using a \emph{rating system} (e.g., Likert scale).
Depending on the underlying task and its criticality, a larger-scale user study must be conducted to reflect various stakeholders and their perspectives.
These surveys must follow the principles of good user \emph{survey design} principles~\cite{lazar_research_2017} and their creation should be independent from a given dataset to ensure an automated reuse for new or changed datasets. 



\section{Vision: A DQ Assessment Framework}
\label{sec:pragmatic}

In previous sections, we explored the challenges associated with different \emph{facets} of DQ assessment and their applications to DQ dimensions.
To promote this fresh look on DQ research, we envision a \textit{DQ assessment framework} that implements the assessment methods along the \emph{facets}. 

\begin{figure}[t]
\includegraphics[width=\linewidth]{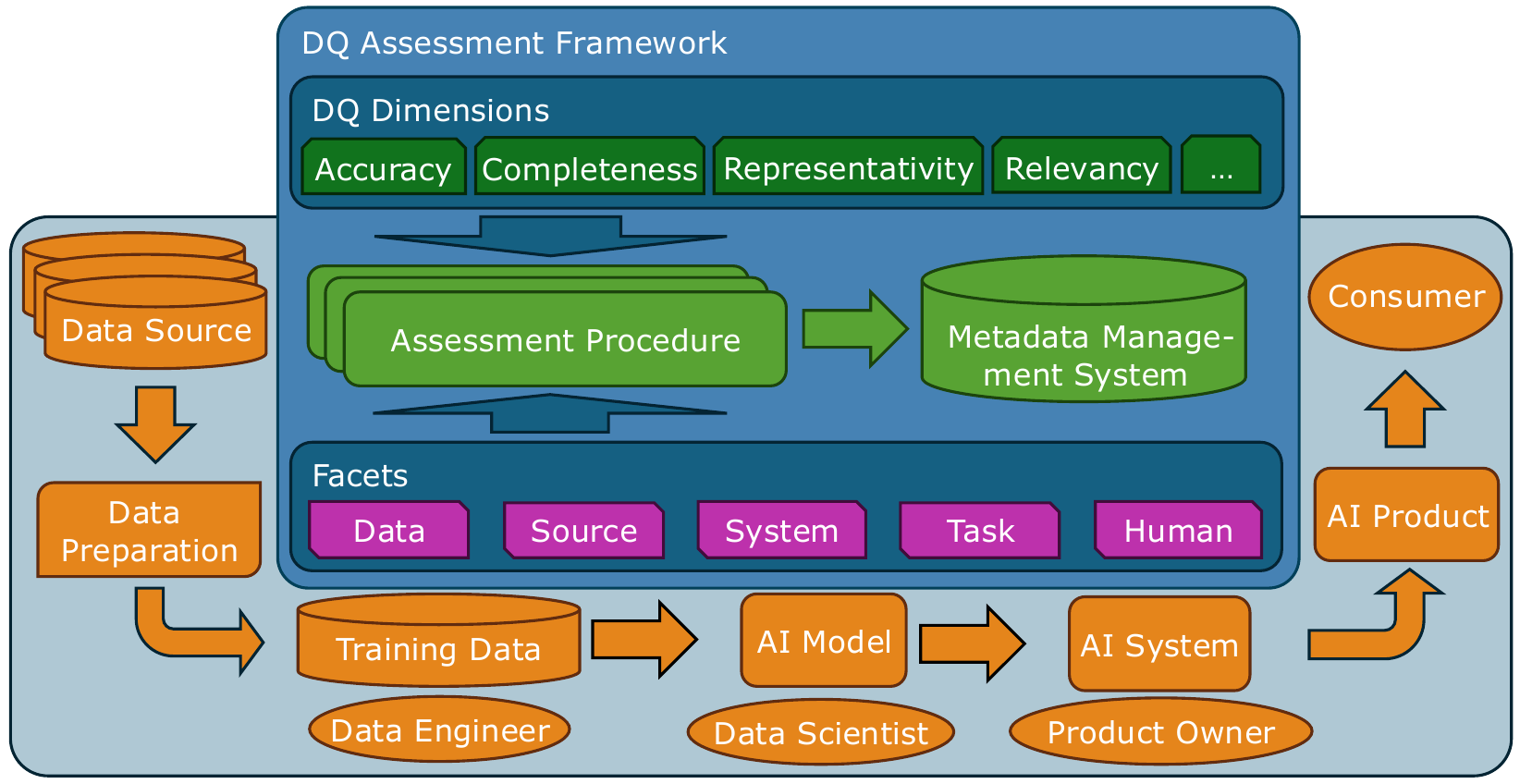}
    \caption{DQ assessment for an AI pipeline.}
    \label{fig:dq_assessment_framework}
\end{figure}

Figure~\ref{fig:dq_assessment_framework} shows the DQ assessment framework in the context of an AI pipeline.
\revision{As part of this pipeline, data passes through various stages from its creation to the final product delivered to the customer.
We can map the facets to these different stages of the pipeline. 
Thus, our proposed framework and the concept of facets are integrated into the AI pipeline:}
The data, in its digital representation (data facet) originate from various sources.
A data engineer must prepare them using data preparation techniques, where all transformations must be traceable (source facet).
The prepared data serve as training data, used by a data scientist to train an AI model, constituting a task (task facet).
All these tasks can be deployed in an AI system~(system facet),
managed by a product owner, which in turn, can be part of an AI product that is delivered to customers. 
Finally, the various involved individuals should also be part of the DQ assessment (human facet).
The assessment of each DQ dimension, together with the \textit{facet's} participation, results in a dedicated assessment procedure.

\revision{Let us assume the assessment of \dq{relevancy} (involving the task and human facets): only the corresponding stages and instances of the AI pipeline (i.e., AI model and persons) are involved here. The assessment follows a procedure tailored to the involved facets, where the assessment results are made available in the \textit{Metadata Management System}~\cite{SrivastavaV07}.}

We conducted an initial analysis of the participation of the \emph{facets} per DQ dimension~\cite{mohammed2024data}.
Apart from the facet-specific challenges to measure DQ in its various dimensions, building a framework that supports DQ measurement and management along the entire pipeline gives rise to further challenges:




\mypara{Efficiency}
The assessment effort and time should be low from a user perspective~\cite{Ballou_2006}.
Data consumers might be unable or unwilling to wait for assessment results, and experts might not have much time to complete questionnaires or help in DQ assessment.

\mypara{Explainability}
Due to their ambiguity~\cite{Jayawardene13}, assessment results must be explainable to consumers.
In addition, the results should be traceable to their root cause, enabling measures to improve quality.

\mypara{Metadata Management}
Deploying the DQ assessment procedure requires an effective mechanism to store and query vast, diverse metadata (see \textit{Metadata Management System} in Figure \ref{fig:dq_assessment_framework}).
An example solution and its challenges are discussed in~\cite{SrivastavaV07}.


\section{Related Work}
\label{sec:related}
This section discusses representative works on DQ assessment and compares them to our fresh look through the lens of \emph{facets}.
Over the last decades, a number of DQ assessment frameworks have been proposed~\cite{DBLP:journals/csur/BatiniCFM09,Cichy_2019}.
For instance, \citet{DBLP:journals/jasis/StviliaGTS07} identified various sources for DQ assessment and distinguished intrinsic, relational, and reputational information quality.
\citet{DBLP:journals/csur/BatiniCFM09} divide the assessment into different phases and discuss metrics for DQ dimensions.
\citet{10.1145/505248.506010} present an approach combining subjective and objective DQ assessment results.
%
%
In their vision paper, Sadiq et al.\ identify two dimensions to empirical DQ management~\cite{Sadiq2018}: the \emph{metric} type (intrinsic vs.\ extrinsic) and the method \emph{scope} (generic vs.\ tailored). They encourage the community to regard DQ beyond what we call the data facet -- this paper follows that call.
Other works~\cite{batini_data_2006,batini_data_2016,10.1145/505248.506010} discuss challenges associated with specific DQ dimensions, e.g., the need for external data to assess \dq{accuracy}~\cite{batini_data_2006}.

In summary, many existing works implicitly mention individual facets (e.g., the human or the data facet) and the impact of their challenges on the assessment of DQ dimensions.
However, so far, a unified view on how to address these different aspects was missing.
We believe that addressing common DQ challenges per \emph{facet} enables researchers the exploration of many DQ dimensions jointly.
\section{Conclusion}
\label{sec:conclusion}

We propose five assessment \emph{facets} as foundational ingredients to assess \textit{data quality} (DQ) and outline specific challenges and opportunities for each \emph{facet}, highlighting the complexity of DQ assessment.
We suggest how to overcome these challenges for the DQ dimensions mentioned in the AI Act as examples.
Finally, we envision a DQ assessment framework that implements various methods to assess the DQ dimension through the lens of the \emph{facets}.
\section*{Acknowledgements}
This research was partially funded by the \href{https://www.kitqar.de/de}{KITQAR} project, supported by Denkfabrik Digitale Arbeitsgemeinschaft im Bundesministerium für Arbeit und Soziales (BMAS). 
\balance



\bibliographystyle{plainnat}
\bibliography{references}

\providecommand{\noopsort}[1]{}
\begin{thebibliography}{81}
\providecommand{\natexlab}[1]{#1}
\providecommand{\url}[1]{\texttt{#1}}
\expandafter\ifx\csname urlstyle\endcsname\relax
  \providecommand{\doi}[1]{doi: #1}\else
  \providecommand{\doi}{doi: \begingroup \urlstyle{rm}\Url}\fi

\bibitem[ai_(2023)]{ai_act_risk_levels}
{EU} {AI} act: first regulation on artificial intelligence, 2023.
\newblock URL
  \url{https://www.europarl.europa.eu/topics/en/article/20230601STO93804/eu-ai-act-first-regulation-on-artificial-intelligence}.
\newblock (Last accessed: 2024-07-25).

\bibitem[noa(2024{\natexlab{a}})]{noauthor_hipaa_2023}
{HIPAA} privacy rule to support reproductive health care privacy,
  2024{\natexlab{a}}.
\newblock URL
  \url{https://www.federalregister.gov/documents/2024/04/26/2024-08503/hipaa-privacy-rule-to-support-reproductive-health-care-privacy}.
\newblock (Last accessed: 2024-07-25).

\bibitem[noa(2024{\natexlab{b}})]{noauthor_kaggle_nodate}
Kaggle: Your machine learning and data science community, 2024{\natexlab{b}}.
\newblock URL \url{https://www.kaggle.com/}.
\newblock (Last accessed: 2024-07-15).

\bibitem[noa(2024{\natexlab{c}})]{noauthor_wikipedia_nodate}
Wikipedia, the free encyclopedia, 2024{\natexlab{c}}.
\newblock URL \url{https://www.wikipedia.org/}.
\newblock (Last accessed: 2024-07-15).

\bibitem[Abedjan et~al.(2015)Abedjan, Golab, and Naumann]{Abedjan15}
Ziawasch Abedjan, Lukasz Golab, and Felix Naumann.
\newblock Profiling relational data: a survey.
\newblock \emph{VLDB Journal}, 24\penalty0 (4):\penalty0 557--581, 2015.
\newblock \doi{10.1007/S00778-015-0389-Y}.

\bibitem[Acar et~al.(2018)Acar, Aksu, Uluagac, and
  Conti]{DBLP:journals/csur/AcarAUC18}
Abbas Acar, Hidayet Aksu, A.~Selcuk Uluagac, and Mauro Conti.
\newblock A survey on homomorphic encryption schemes: Theory and
  implementation.
\newblock \emph{ACM Computing Surveys}, 51\penalty0 (4):\penalty0 79:1--79:35,
  2018.
\newblock \doi{10.1145/3214303}.
\newblock URL \url{https://doi.org/10.1145/3214303}.

\bibitem[Akhtar et~al.(2024)Akhtar, Benjelloun, Conforti, Gijsbers,
  Giner{-}Miguelez, Jain, Kuchnik, Lhoest, Marcenac, Maskey, Mattson, Oala,
  Ruyssen, Shinde, Simperl, Thomas, Tykhonov, Vanschoren, van~der Velde,
  Vogler, and Wu]{akhtar2024croissant}
Mubashara Akhtar, Omar Benjelloun, Costanza Conforti, Pieter Gijsbers, Joan
  Giner{-}Miguelez, Nitisha Jain, Michael Kuchnik, Quentin Lhoest, Pierre
  Marcenac, Manil Maskey, Peter Mattson, Luis Oala, Pierre Ruyssen, Rajat
  Shinde, Elena Simperl, Goeffry Thomas, Slava Tykhonov, Joaquin Vanschoren,
  Jos van~der Velde, Steffen Vogler, and Carole{-}Jean Wu.
\newblock Croissant: {A} metadata format for ml-ready datasets.
\newblock In \emph{Proceedings of the International Conference on Management of
  Data (SIGMOD)}, pages 1--6. {ACM}, 2024.
\newblock \doi{10.1145/3650203.3663326}.
\newblock URL \url{https://doi.org/10.1145/3650203.3663326}.

\bibitem[Asudeh et~al.(2019)Asudeh, Jin, and Jagadish]{asudeh2019assessing}
Abolfazl Asudeh, Zhongjun Jin, and HV~Jagadish.
\newblock Assessing and remedying coverage for a given dataset.
\newblock In \emph{2019 IEEE 35th International Conference on Data Engineering
  (ICDE)}, pages 554--565. IEEE, 2019.

\bibitem[Ballou et~al.(2006)Ballou, Chengalur-Smith, and Wang]{Ballou_2006}
Donald~P Ballou, InduShobha~N Chengalur-Smith, and Richard~Y Wang.
\newblock Sample-based quality estimation of query results in relational
  database environments.
\newblock \emph{IEEE Transactions on Knowledge and Data Engineering},
  18\penalty0 (5):\penalty0 639--650, 2006.

\bibitem[Batini and Scannapieco(2006)]{batini_data_2006}
Carlo Batini and Monica Scannapieco.
\newblock \emph{Data quality: concepts, methodologies and techniques}.
\newblock Data-centric systems and applications. Springer, 2006.
\newblock ISBN 978-3-540-33172-8 978-3-642-06970-3.

\bibitem[Batini and Scannapieco(2016)]{batini_data_2016}
Carlo Batini and Monica Scannapieco.
\newblock \emph{Data and Information Quality: Dimensions, Principles and
  Techniques}.
\newblock Springer Berlin Heidelberg, 2016.
\newblock ISBN 978-3-319-24104-3.

\bibitem[Batini et~al.(1986)Batini, Lenzerini, and
  Navathe]{DBLP:journals/csur/BatiniL86}
Carlo Batini, Maurizio Lenzerini, and Shamkant~B. Navathe.
\newblock A comparative analysis of methodologies for database schema
  integration.
\newblock \emph{ACM Computing Surveys}, 18\penalty0 (4):\penalty0 323--364,
  1986.
\newblock \doi{10.1145/27633.27634}.
\newblock URL \url{https://doi.org/10.1145/27633.27634}.

\bibitem[Batini et~al.(2009)Batini, Cappiello, Francalanci, and
  Maurino]{DBLP:journals/csur/BatiniCFM09}
Carlo Batini, Cinzia Cappiello, Chiara Francalanci, and Andrea Maurino.
\newblock Methodologies for data quality assessment and improvement.
\newblock \emph{ACM Computing Surveys}, 41\penalty0 (3):\penalty0 16:1--16:52,
  2009.
\newblock \doi{10.1145/1541880.1541883}.
\newblock URL \url{https://doi.org/10.1145/1541880.1541883}.

\bibitem[Bechny et~al.(2021)Bechny, Sobieczky, Zeindl, and
  Ehrlinger]{Bechny_2021}
Michal Bechny, Florian Sobieczky, J\"{u}rgen Zeindl, and Lisa Ehrlinger.
\newblock Missing data patterns: From theory to an application in the steel
  industry.
\newblock In \emph{Proceedings of the International Conference on Scientific
  and Statistical Database Management (SSDBM)}, page 214–219, New York, NY,
  USA, 2021. Association for Computing Machinery.
\newblock ISBN 9781450384131.
\newblock \doi{10.1145/3468791.3468841}.
\newblock URL \url{https://doi.org/10.1145/3468791.3468841}.

\bibitem[Berendt et~al.(2002)Berendt, Hotho, and
  Stumme]{10.1007/3-540-48005-6_21}
Bettina Berendt, Andreas Hotho, and Gerd Stumme.
\newblock Towards semantic web mining.
\newblock In Ian Horrocks and James~A. Hendler, editors, \emph{Proceedings of
  the International Semantic Web Conference (ISWC)}, volume 2342 of
  \emph{Lecture Notes in Computer Science}, pages 264--278. Springer, 2002.
\newblock \doi{10.1007/3-540-48005-6\_21}.
\newblock URL \url{https://doi.org/10.1007/3-540-48005-6\_21}.

\bibitem[Bertino et~al.(2019)Bertino, Kundu, and
  Sura]{DBLP:journals/jdiq/BertinoKS19}
Elisa Bertino, Ashish Kundu, and Zehra Sura.
\newblock Data transparency with blockchain and {AI} ethics.
\newblock \emph{Journal on Data and Information Quality}, 11\penalty0
  (4):\penalty0 16:1--16:8, 2019.
\newblock \doi{10.1145/3312750}.
\newblock URL \url{https://doi.org/10.1145/3312750}.

\bibitem[Budach et~al.(2022)Budach, Feuerpfeil, Ihde, Nathansen, Noack,
  Patzlaff, Harmouch, and Naumann]{budach2022effects}
Lukas Budach, Moritz Feuerpfeil, Nina Ihde, Andrea Nathansen, Nele Noack,
  Hendrik Patzlaff, Hazar Harmouch, and Felix Naumann.
\newblock The effects of data quality on machine learning performance.
\newblock \emph{arXiv preprint arXiv:2207.14529}, 2022.
\newblock URL \url{https://doi.org/10.48550/arXiv.2207.14529}.

\bibitem[Budka et~al.(2011)Budka, Gabrys, and
  Musial]{DBLP:journals/entropy/BudkaGM11}
Marcin Budka, Bogdan Gabrys, and Katarzyna Musial.
\newblock On accuracy of {PDF} divergence estimators and their applicability to
  representative data sampling.
\newblock \emph{Entropy}, 13\penalty0 (7):\penalty0 1229--1266, 2011.
\newblock \doi{10.3390/E13071229}.
\newblock URL \url{https://doi.org/10.3390/e13071229}.

\bibitem[Cai and Zhu(2015)]{DBLP:journals/datascience/CaiZ15}
Li~Cai and Yangyong Zhu.
\newblock The challenges of data quality and data quality assessment in the big
  data era.
\newblock \emph{Data Sci. J.}, 14:\penalty0 2, 2015.
\newblock \doi{10.5334/DSJ-2015-002}.
\newblock URL \url{https://doi.org/10.5334/dsj-2015-002}.

\bibitem[Christen(2012)]{Christen12}
Peter Christen.
\newblock \emph{Data Matching - Concepts and Techniques for Record Linkage,
  Entity Resolution, and Duplicate Detection}.
\newblock Data-Centric Systems and Applications. Springer, 2012.
\newblock \doi{10.1007/978-3-642-31164-2}.
\newblock URL \url{https://doi.org/10.1007/978-3-642-31164-2}.

\bibitem[Cichy and Rass(2019)]{Cichy_2019}
Corinna Cichy and Stefan Rass.
\newblock An overview of data quality frameworks.
\newblock \emph{IEEE Access}, 7:\penalty0 24634--24648, 2019.

\bibitem[Clemmensen and
  Kj{\ae}rsgaard(2022)]{DBLP:journals/corr/abs-2203-04706}
Line~H. Clemmensen and Rune~D. Kj{\ae}rsgaard.
\newblock Data representativity for machine learning and {AI} systems.
\newblock \emph{CoRR}, abs/2203.04706, 2022.
\newblock \doi{10.48550/ARXIV.2203.04706}.
\newblock URL \url{https://doi.org/10.48550/arXiv.2203.04706}.

\bibitem[Dallachiesa et~al.(2013)Dallachiesa, Ebaid, Eldawy, Elmagarmid, Ilyas,
  Ouzzani, and Tang]{DBLP:conf/sigmod/DallachiesaEEEIOT13}
Michele Dallachiesa, Amr Ebaid, Ahmed Eldawy, Ahmed~K. Elmagarmid, Ihab~F.
  Ilyas, Mourad Ouzzani, and Nan Tang.
\newblock {NADEEF:} a commodity data cleaning system.
\newblock In \emph{Proceedings of the International Conference on Management of
  Data (SIGMOD)}, pages 541--552. {ACM}, 2013.
\newblock \doi{10.1145/2463676.2465327}.
\newblock URL \url{https://doi.org/10.1145/2463676.2465327}.

\bibitem[Doan et~al.(2012)Doan, Halevy, and Ives]{doan2012principles}
AnHai Doan, Alon Halevy, and Zachary Ives.
\newblock \emph{Principles of Data Integration}.
\newblock Morgan Kaufmann, 2012.
\newblock ISBN 978-0-12-416044-6.
\newblock \doi{10.1016/C2011-0-06130-6}.
\newblock URL \url{https://doi.org/10.1016/C2011-0-06130-6}.

\bibitem[Dwork(2006)]{hutchison_differential_2006}
Cynthia Dwork.
\newblock Differential privacy.
\newblock In \emph{Automata, Languages and Programming}, volume 4052, pages
  1--12. Springer Berlin Heidelberg, 2006.
\newblock ISBN 978-3-540-35907-4 978-3-540-35908-1.
\newblock \doi{10.1007/11787006_1}.
\newblock URL \url{http://link.springer.com/10.1007/11787006_1}.
\newblock Series Title: Lecture Notes in Computer Science.

\bibitem[Ehrlinger et~al.(2021)Ehrlinger, Schrott, Melichar, Kirchmayr, and
  W{\"{o}}{\ss}]{DBLP:conf/dexaw/EhrlingerSMKW21}
Lisa Ehrlinger, Johannes Schrott, Martin Melichar, Nicolas Kirchmayr, and
  Wolfram W{\"{o}}{\ss}.
\newblock Data catalogs: {A} systematic literature review and guidelines to
  implementation.
\newblock In \emph{DEXA Workshops Proceedings}, volume 1479 of
  \emph{Communications in Computer and Information Science}, pages 148--158.
  Springer, 2021.
\newblock \doi{10.1007/978-3-030-87101-7\_15}.
\newblock URL \url{https://doi.org/10.1007/978-3-030-87101-7\_15}.

\bibitem[Ehrlinger et~al.(2023)Ehrlinger, Schrott, and W{\"o}{\ss}]{dsd_2023}
Lisa Ehrlinger, Johannes Schrott, and Wolfram W{\"o}{\ss}.
\newblock Dsd: the data source description vocabulary.
\newblock In \emph{International Conference on Database and Expert Systems
  Applications (DEXA)}, pages 3--10. Springer, 2023.

\bibitem[{European Parliament}(2024)]{european_parliament_artifical_2024}
{European Parliament}.
\newblock Artifical inteligence act.
\newblock 2024.
\newblock URL
  \url{https://eur-lex.europa.eu/legal-content/EN/TXT/?uri=CELEX:32024R1689}.
\newblock {Version from 2024-06-13}.

\bibitem[Fernandez et~al.(2023)Fernandez, Elmore, Franklin, Krishnan, and
  Tan]{DBLP:journals/pvldb/FernandezEFKT23}
Raul~Castro Fernandez, Aaron~J. Elmore, Michael~J. Franklin, Sanjay Krishnan,
  and Chenhao Tan.
\newblock How large language models will disrupt data management.
\newblock \emph{PVLDB}, 16\penalty0 (11):\penalty0 3302--3309, 2023.
\newblock \doi{10.14778/3611479.3611527}.
\newblock URL \url{https://www.vldb.org/pvldb/vol16/p3302-fernandez.pdf}.

\bibitem[for Standardization(2015)]{ISO25024}
International~Organization for Standardization.
\newblock Iso/iec 25024:2015 systems and software engineering -- systems and
  software quality requirements and evaluation (square) -- measurement of data
  quality.
\newblock Technical report, International Organization for Standardization,
  2015.
\newblock URL \url{https://www.iso.org/standard/35762.html}.
\newblock ISO/IEC 25024:2015.

\bibitem[Foroni et~al.(2021)Foroni, Lissandrini, and
  Velegrakis]{DBLP:conf/icde/ForoniLV21}
Daniele Foroni, Matteo Lissandrini, and Yannis Velegrakis.
\newblock Estimating the extent of the effects of data quality through
  observations.
\newblock In \emph{Proceedings of the International Conference on Data
  Engineering (ICDE)}, pages 1913--1918. {IEEE}, 2021.
\newblock \doi{10.1109/ICDE51399.2021.00176}.
\newblock URL \url{https://doi.org/10.1109/ICDE51399.2021.00176}.

\bibitem[GDPR()]{General_Data_Protection_Regulation}
GDPR.
\newblock General data protection regulation (last accessed: 2024-02-13), 2016.
\newblock URL
  \url{https://eur-lex.europa.eu/legal-content/EN/TXT/PDF/?uri=CELEX:02016R0679-20160504}.

\bibitem[Glavic and Dittrich(2007)]{DBLP:conf/btw/GlavicD07}
Boris Glavic and Klaus~R. Dittrich.
\newblock Data provenance: {A} categorization of existing approaches.
\newblock In \emph{Proceedings of the Conference Datenbanksysteme in Business,
  Technologie und Web Technik (BTW)}, volume {P-103} of \emph{{LNI}}, pages
  227--241. {GI}, 2007.
\newblock URL \url{https://dl.gi.de/handle/20.500.12116/31801}.

\bibitem[Grossman and Frieder(2004)]{grossman_information_2004}
David~A. Grossman and Ophir Frieder.
\newblock \emph{Information retrieval: algorithms and heuristics}.
\newblock Number~15. Springer, 2nd ed edition, 2004.
\newblock ISBN 978-1-4020-3004-8 978-1-4020-3003-1.

\bibitem[Haegemans et~al.(2016)Haegemans, Snoeck, and Lemahieu]{Haegemans_2016}
Tom Haegemans, Monique Snoeck, and Wilfried Lemahieu.
\newblock Towards a precise definition of data accuracy and a justification for
  its measure.
\newblock In \emph{Proceedings of the International Conference on Information
  Quality}, pages 16--16. MIT Information Quality (MITIQ) Program, 2016.

\bibitem[He and Garcia(2009)]{DBLP:journals/tkde/HeG09}
Haibo He and Edwardo~A. Garcia.
\newblock Learning from imbalanced data.
\newblock \emph{IEEE Transactions on Knowledge and Data Engineering (TKDE)},
  21\penalty0 (9):\penalty0 1263--1284, 2009.
\newblock \doi{10.1109/TKDE.2008.239}.
\newblock URL \url{https://doi.org/10.1109/TKDE.2008.239}.

\bibitem[Heise et~al.(2014)Heise, Kasneci, and Naumann]{Heise14}
Arvid Heise, Gjergji Kasneci, and Felix Naumann.
\newblock Estimating the number and sizes of fuzzy-duplicate clusters.
\newblock In \emph{Proceedings of the International Conference on Information
  and Knowledge Management (CIKM)}, pages 959--968. {ACM}, 2014.
\newblock \doi{10.1145/2661829.2661885}.
\newblock URL \url{https://doi.org/10.1145/2661829.2661885}.

\bibitem[Herschel et~al.(2017)Herschel, Diestelk{\"{a}}mper, and {Ben
  Lahmar}]{DBLP:journals/vldb/HerschelDL17}
Melanie Herschel, Ralf Diestelk{\"{a}}mper, and Houssem {Ben Lahmar}.
\newblock A survey on provenance: What for? what form? what from?
\newblock \emph{VLDB Journal}, 26\penalty0 (6):\penalty0 881--906, 2017.
\newblock \doi{10.1007/S00778-017-0486-1}.
\newblock URL \url{https://doi.org/10.1007/s00778-017-0486-1}.

\bibitem[Herzog et~al.(2007)Herzog, Scheuren, and Winkler]{herzog_data_2007}
Thomas~N. Herzog, Fritz Scheuren, and William~E. Winkler.
\newblock \emph{Data quality and record linkage techniques}.
\newblock Springer, 2007.
\newblock ISBN 978-0-387-69502-0.
\newblock {OCLC}: ocn137313060.

\bibitem[House(2023)]{house_executive_2023}
The~White House.
\newblock Executive order on the safe, secure, and trustworthy development and
  use of artificial intelligence, 2023.
\newblock URL
  \url{https://www.whitehouse.gov/briefing-room/presidential-actions/2023/10/30/executive-order-on-the-safe-secure-and-trustworthy-development-and-use-of-artificial-intelligence/}.

\bibitem[Jayawardene et~al.(2013)Jayawardene, Sadiq, and
  Indulska]{Jayawardene13}
Vimukthi Jayawardene, Shazia~W. Sadiq, and Marta Indulska.
\newblock The curse of dimensionality in data quality.
\newblock In \emph{Australasian Conference on Information Systems (ACIS)}, page
  165, 2013.
\newblock URL \url{https://aisel.aisnet.org/acis2013/165}.

\bibitem[Khayyat et~al.(2015)Khayyat, Ilyas, Jindal, Madden, Ouzzani, Papotti,
  Quian{\'{e}}{-}Ruiz, Tang, and Yin]{KhayyatDansing15}
Zuhair Khayyat, Ihab~F. Ilyas, Alekh Jindal, Samuel Madden, Mourad Ouzzani,
  Paolo Papotti, Jorge{-}Arnulfo Quian{\'{e}}{-}Ruiz, Nan Tang, and Si~Yin.
\newblock Bigdansing: {A} system for big data cleansing.
\newblock In \emph{Proceedings of the International Conference on Management of
  Data (SIGMOD)}, pages 1215--1230. {ACM}, 2015.
\newblock \doi{10.1145/2723372.2747646}.
\newblock URL \url{https://doi.org/10.1145/2723372.2747646}.

\bibitem[Kruskal and Mosteller(1979)]{10bd9b32-8b48-36d1-a9cc-51fbb4bf7d53}
William Kruskal and Frederick Mosteller.
\newblock Representative sampling, {III}: The current statistical literature.
\newblock \emph{International Statistical Review / Revue Internationale de
  Statistique}, 47\penalty0 (3):\penalty0 245--265, 1979.
\newblock \doi{10.2307/1402647}.

\bibitem[Kuebler-Wachendorff et~al.(2021)Kuebler-Wachendorff, Luzsa, Kranz,
  Mager, Syrmoudis, Mayr, and Grossklags]{kuebler-wachendorff_right_2021}
Sophie Kuebler-Wachendorff, Robert Luzsa, Johann Kranz, Stefan Mager, Emmanuel
  Syrmoudis, Susanne Mayr, and Jens Grossklags.
\newblock The right to data portability: conception, status quo, and future
  directions.
\newblock \emph{Informatik Spektrum}, 44\penalty0 (4):\penalty0 264--272, 2021.
\newblock ISSN 0170-6012, 1432-122X.
\newblock \doi{10.1007/s00287-021-01372-w}.
\newblock URL \url{https://link.springer.com/10.1007/s00287-021-01372-w}.

\bibitem[Lazar et~al.(2017)Lazar, Feng, and Hochheiser]{lazar_research_2017}
Jonathan Lazar, Jinjuan~Heidi Feng, and Harry Hochheiser.
\newblock \emph{Research Methods in Human Computer Interaction}.
\newblock Elsevier, second edition, 2017.
\newblock ISBN 978-0-12-805390-4.

\bibitem[Lehmann et~al.(2015)Lehmann, Isele, Jakob, Jentzsch, Kontokostas,
  Mendes, Hellmann, Morsey, van Kleef, Auer, and Bizer]{DBpedia15}
Jens Lehmann, Robert Isele, Max Jakob, Anja Jentzsch, Dimitris Kontokostas,
  Pablo~N. Mendes, Sebastian Hellmann, Mohamed Morsey, Patrick van Kleef,
  S{\"{o}}ren Auer, and Christian Bizer.
\newblock {DBpedia} - {A} large-scale, multilingual knowledge base extracted
  from wikipedia.
\newblock \emph{Semantic Web}, 6\penalty0 (2):\penalty0 167--195, 2015.
\newblock \doi{10.3233/SW-140134}.

\bibitem[Li et~al.(2021)Li, Rao, Blase, Zhang, Chu, and
  Zhang]{DBLP:conf/icde/LiRBZCZ21}
Peng Li, Xi~Rao, Jennifer Blase, Yue Zhang, Xu~Chu, and Ce~Zhang.
\newblock Cleanml: {A} study for evaluating the impact of data cleaning on {ML}
  classification tasks.
\newblock In \emph{Proceedings of the International Conference on Data
  Engineering (ICDE)}, pages 13--24. {IEEE}, 2021.
\newblock \doi{10.1109/ICDE51399.2021.00009}.
\newblock URL \url{https://doi.org/10.1109/ICDE51399.2021.00009}.

\bibitem[Magurran(2021)]{magurran_measuring_2021}
Anne~E. Magurran.
\newblock Measuring biological diversity.
\newblock \emph{Current Biology}, 31\penalty0 (19):\penalty0 R1174--R1177,
  2021.
\newblock ISSN 09609822.
\newblock \doi{10.1016/j.cub.2021.07.049}.
\newblock URL
  \url{https://linkinghub.elsevier.com/retrieve/pii/S0960982221010393}.

\bibitem[Maydanchik(2007)]{maydanchik_data_2007}
Arkady Maydanchik.
\newblock \emph{Data quality assessment}.
\newblock Data quality for practitioners series. Technics Publications, 2007.
\newblock ISBN 978-0-9771400-2-2.

\bibitem[Milo and Zohar(1998)]{DBLP:conf/vldb/MiloZ98}
Tova Milo and Sagit Zohar.
\newblock Using schema matching to simplify heterogeneous data translation.
\newblock In \emph{Proceedings of the International Conference on Very Large
  Databases (VLDB)}, pages 122--133, 1998.
\newblock URL \url{http://www.vldb.org/conf/1998/p122.pdf}.

\bibitem[Mohammed et~al.(2024{\natexlab{a}})Mohammed, Brandner, Burtscher,
  Hallensleben, Harmouch, Hauschke, Heesen, Hildebrandt, Hirsbrunner, Keselj,
  Mahlow, Massow, Naumann, Rostalski, Wilken, and
  Wölke]{mohammed_glossary_2024}
Sedir Mohammed, Lou~Therese Brandner, Felicia Burtscher, Sebastian
  Hallensleben, Hazar Harmouch, Andreas Hauschke, Jessica Heesen, Stefanie
  Hildebrandt, Simon~David Hirsbrunner, Julia Keselj, Philipp Mahlow, Marie
  Massow, Felix Naumann, Frauke Rostalski, Anna Wilken, and Annika Wölke.
\newblock A data quality glossary.
\newblock 2024{\natexlab{a}}.
\newblock \doi{10.5281/ZENODO.10474880}.
\newblock URL \url{https://zenodo.org/doi/10.5281/zenodo.10474880}.

\bibitem[Mohammed et~al.(2024{\natexlab{b}})Mohammed, Harmouch, Naumann, and
  Srivastava]{mohammed2024data}
Sedir Mohammed, Hazar Harmouch, Felix Naumann, and Divesh Srivastava.
\newblock Data quality assessment: Challenges and opportunities.
\newblock \emph{CoRR}, abs/2403.00526, 2024{\natexlab{b}}.
\newblock \doi{10.48550/ARXIV.2403.00526}.
\newblock URL \url{https://doi.org/10.48550/arXiv.2403.00526}.

\bibitem[Moraga et~al.(2009)Moraga, Moraga, Calero, and
  Caro]{DBLP:conf/qsic/MoragaMCC09}
Carmen Moraga, Mar{\'{\i}}a~{\'{A}}ngeles Moraga, Coral Calero, and
  Ang{\'{e}}lica Caro.
\newblock Square-aligned data quality model for web portals.
\newblock In \emph{Proceedings of the Ninth International Conference on Quality
  Software (QSIC)}, pages 117--122. {IEEE}, 2009.
\newblock \doi{10.1109/QSIC.2009.23}.
\newblock URL \url{https://doi.org/10.1109/QSIC.2009.23}.

\bibitem[Nagle et~al.(2020)Nagle, Redman, and Sammon]{nagle_assessing_2020}
Tadhg Nagle, Tom Redman, and David Sammon.
\newblock Assessing data quality: A managerial call to action.
\newblock \emph{Business Horizons}, 63\penalty0 (3):\penalty0 325--337, 2020.
\newblock ISSN 00076813.
\newblock \doi{10.1016/j.bushor.2020.01.006}.
\newblock URL
  \url{https://linkinghub.elsevier.com/retrieve/pii/S0007681320300069}.

\bibitem[Naumann(2013)]{DBLP:journals/sigmod/Naumann13}
Felix Naumann.
\newblock Data profiling revisited.
\newblock \emph{{SIGMOD} Rec.}, 42\penalty0 (4):\penalty0 40--49, 2013.
\newblock \doi{10.1145/2590989.2590995}.
\newblock URL \url{https://doi.org/10.1145/2590989.2590995}.

\bibitem[Naumann and Herschel(2010)]{naumann_introduction_2010}
Felix Naumann and Melanie Herschel.
\newblock \emph{An introduction to duplicate detection}.
\newblock Number~3 in Synthesis lectures on data management. Morgan \& Claypool
  Publishers, 2010.
\newblock ISBN 978-1-60845-220-0.

\bibitem[Naumann and Rolker(2000)]{DBLP:conf/iq/NaumannR00}
Felix Naumann and Claudia Rolker.
\newblock Assessment methods for information quality criteria.
\newblock In \emph{Fifth Conference on Information Quality {(IQ} 2000)}, pages
  148--162. {MIT}, 2000.

\bibitem[Neutatz et~al.(2021)Neutatz, Chen, Abedjan, and
  Wu]{neutatz2021cleaning}
Felix Neutatz, Binger Chen, Ziawasch Abedjan, and Eugene Wu.
\newblock From cleaning before {ML} to cleaning for {ML}.
\newblock \emph{IEEE Data Engineering Bulletin}, 44\penalty0 (1):\penalty0
  24--41, 2021.
\newblock URL \url{http://sites.computer.org/debull/A21mar/p24.pdf}.

\bibitem[Neutatz et~al.(2022)Neutatz, Chen, Alkhatib, Ye, and
  Abedjan]{neutatz2022data}
Felix Neutatz, Binger Chen, Yazan Alkhatib, Jingwen Ye, and Ziawasch Abedjan.
\newblock Data cleaning and automl: Would an optimizer choose to clean?
\newblock \emph{Datenbank-Spektrum}, 22\penalty0 (2):\penalty0 121--130, 2022.
\newblock \doi{10.1007/s13222-022-00413-2}.
\newblock URL \url{https://doi.org/10.1007/s13222-022-00413-2}.

\bibitem[Pena et~al.(2020)Pena, Filho, de~Almeida, and Naumann]{PenaVio20}
Eduardo H.~M. Pena, Edson Ramiro~Lucas Filho, Eduardo~C. de~Almeida, and Felix
  Naumann.
\newblock Efficient detection of data dependency violations.
\newblock In \emph{Proceedings of the International Conference on Information
  and Knowledge Management (CIKM)}, pages 1235--1244. {ACM}, 2020.
\newblock \doi{10.1145/3340531.3412062}.
\newblock URL \url{https://doi.org/10.1145/3340531.3412062}.

\bibitem[Pipino et~al.(2002)Pipino, Lee, and Wang]{10.1145/505248.506010}
Leo~L. Pipino, Yang~W. Lee, and Richard~Y. Wang.
\newblock Data quality assessment.
\newblock \emph{Communications of the ACM}, 45\penalty0 (4):\penalty0
  211–218, 2002.
\newblock ISSN 0001-0782.
\newblock \doi{10.1145/505248.506010}.
\newblock URL \url{https://doi.org/10.1145/505248.506010}.

\bibitem[Pushkarna et~al.(2022)Pushkarna, Zaldivar, and
  Kjartansson]{DataCards22}
Mahima Pushkarna, Andrew Zaldivar, and Oddur Kjartansson.
\newblock Data cards: Purposeful and transparent dataset documentation for
  responsible {AI}.
\newblock In \emph{Proceedings of the ACM Conference on Fairness,
  Accountability, and Transparency (FaCCT)}, page 1776–1826, New York, NY,
  USA, 2022. Association for Computing Machinery.
\newblock \doi{10.1145/3531146.3533231}.
\newblock URL \url{https://doi.org/10.1145/3531146.3533231}.

\bibitem[Qahtan et~al.(2018)Qahtan, Elmagarmid, Castro~Fernandez, Ouzzani, and
  Tang]{FAHES2018}
Abdulhakim~A. Qahtan, Ahmed Elmagarmid, Raul Castro~Fernandez, Mourad Ouzzani,
  and Nan Tang.
\newblock Fahes: A robust disguised missing values detector.
\newblock In \emph{Proceedings of the International Conference on Knowledge
  discovery and data mining (SIGKDD)}, page 2100–2109, New York, NY, USA,
  2018. Association for Computing Machinery.
\newblock ISBN 9781450355520.
\newblock \doi{10.1145/3219819.3220109}.
\newblock URL \url{https://doi.org/10.1145/3219819.3220109}.

\bibitem[Rahm and Bernstein(2001)]{DBLP:journals/vldb/RahmB01}
Erhard Rahm and Philip~A. Bernstein.
\newblock A survey of approaches to automatic schema matching.
\newblock \emph{VLDB Journal}, 10\penalty0 (4):\penalty0 334--350, 2001.
\newblock \doi{10.1007/S007780100057}.
\newblock URL \url{https://doi.org/10.1007/s007780100057}.

\bibitem[Redman(2001)]{redman2001data}
Thomas~C Redman.
\newblock \emph{Data quality: the field guide}.
\newblock Digital press, 2001.

\bibitem[Rekatsinas et~al.(2017)Rekatsinas, Chu, Ilyas, and
  R{\'{e}}]{DBLP:journals/corr/RekatsinasCIR17}
Theodoros Rekatsinas, Xu~Chu, Ihab~F. Ilyas, and Christopher R{\'{e}}.
\newblock {HoloClean}: Holistic data repairs with probabilistic inference.
\newblock \emph{PVLDB}, 10\penalty0 (11):\penalty0 1190--1201, 2017.
\newblock \doi{10.14778/3137628.3137631}.
\newblock URL \url{http://www.vldb.org/pvldb/vol10/p1190-rekatsinas.pdf}.

\bibitem[Roberts et~al.(2021)Roberts, Cowls, Morley, Taddeo, Wang, and
  Floridi]{roberts_chinese_2021}
Huw Roberts, Josh Cowls, Jessica Morley, Mariarosaria Taddeo, Vincent Wang, and
  Luciano Floridi.
\newblock The chinese approach to artificial intelligence: an analysis of
  policy, ethics, and regulation.
\newblock \emph{{AI} \& {SOCIETY}}, 36\penalty0 (1):\penalty0 59--77, 2021.
\newblock ISSN 0951-5666, 1435-5655.
\newblock \doi{10.1007/s00146-020-00992-2}.
\newblock URL \url{https://link.springer.com/10.1007/s00146-020-00992-2}.

\bibitem[Sadiq(2013)]{sadiq_handbook_2013}
Shazia Sadiq, editor.
\newblock \emph{Handbook of data quality: research and practice}.
\newblock Springer, 2013.
\newblock ISBN 978-3-642-36256-9.
\newblock \doi{10.1007/978-3-642-36257-6}.

\bibitem[Sadiq et~al.(2018)Sadiq, Dasu, Dong, Freire, Ilyas, Link, Miller,
  Naumann, Zhou, and Srivastava]{Sadiq2018}
Shazia Sadiq, Tamraparni Dasu, Xin~Luna Dong, Juliana Freire, Ihab~F. Ilyas,
  Sebastian Link, Miller~J. Miller, Felix Naumann, Xiaofang Zhou, and Divesh
  Srivastava.
\newblock Data quality: The role of empiricism.
\newblock \emph{SIGMOD Record}, 46\penalty0 (4):\penalty0 35–43, 2018.
\newblock URL \url{https://doi.org/10.1145/3186549.3186559}.

\bibitem[Shah et~al.(2002)Shah, Finin, and Joshi]{DBLP:conf/cikm/ShahFJ02}
Urvi Shah, Timothy~W. Finin, and Anupam Joshi.
\newblock Information retrieval on the semantic web.
\newblock In \emph{Proceedings of the International Conference on Information
  and Knowledge Management (CIKM)}, pages 461--468, 2002.
\newblock \doi{10.1145/584792.584868}.
\newblock URL \url{https://doi.org/10.1145/584792.584868}.

\bibitem[Shah et~al.(2024)Shah, Parashos, and Kumar]{shah_how_2024}
Vraj Shah, Thomas Parashos, and Arun Kumar.
\newblock How do categorical duplicates affect {ML}? a new benchmark and
  empirical analyses.
\newblock Technical report, 2024.
\newblock URL \url{https://adalabucsd.github.io/papers/TR_2023_CategDedup.pdf}.

\bibitem[Shapley(1953)]{shapley1953value}
Lloyd~S Shapley.
\newblock A value for n-person games.
\newblock In \emph{Contributions to the Theory of Games II}, pages 307--317.
  Princeton University Press, Princeton, 1953.

\bibitem[Slack et~al.(2021)Slack, Hilgard, Singh, and
  Lakkaraju]{DBLP:conf/nips/SlackHSL21}
Dylan Slack, Anna Hilgard, Sameer Singh, and Himabindu Lakkaraju.
\newblock Reliable post hoc explanations: Modeling uncertainty in
  explainability.
\newblock In \emph{Annual Conference on Neural Information Processing Systems
  (NeurIPS)}, pages 9391--9404, 2021.
\newblock URL
  \url{https://proceedings.neurips.cc/paper/2021/hash/4e246a381baf2ce038b3b0f82c7d6fb4-Abstract.html}.

\bibitem[Srivastava and Velegrakis(2007)]{SrivastavaV07}
Divesh Srivastava and Yannis Velegrakis.
\newblock Intensional associations between data and metadata.
\newblock In \emph{Proceedings of the International Conference on Management of
  Data (SIGMOD)}, pages 401--412. {ACM}, 2007.
\newblock \doi{10.1145/1247480.1247526}.

\bibitem[Stodden(2020)]{Stodden20}
Victoria Stodden.
\newblock The data science life cycle: a disciplined approach to advancing data
  science as a science.
\newblock \emph{Communications of the ACM}, 63\penalty0 (7):\penalty0 58--66,
  2020.
\newblock \doi{10.1145/3360646}.
\newblock URL \url{https://doi.org/10.1145/3360646}.

\bibitem[Stvilia et~al.(2007)Stvilia, Gasser, Twidale, and
  Smith]{DBLP:journals/jasis/StviliaGTS07}
Besiki Stvilia, Les Gasser, Michael~B. Twidale, and Linda~C. Smith.
\newblock A framework for information quality assessment.
\newblock \emph{J. Assoc. Inf. Sci. Technol.}, 58\penalty0 (12):\penalty0
  1720--1733, 2007.
\newblock \doi{10.1002/ASI.20652}.
\newblock URL \url{https://doi.org/10.1002/asi.20652}.

\bibitem[Sundararajan and Najmi(2020)]{DBLP:conf/icml/SundararajanN20}
Mukund Sundararajan and Amir Najmi.
\newblock The many {Shapley} values for model explanation.
\newblock In \emph{Proceedings of the International Conference on Machine
  Learning (ICML)}, volume 119, pages 9269--9278. {PMLR}, 2020.
\newblock URL \url{http://proceedings.mlr.press/v119/sundararajan20b.html}.

\bibitem[Sweeney(2002)]{DBLP:journals/ijufks/Sweene02}
Latanya Sweeney.
\newblock k-anonymity: {A} model for protecting privacy.
\newblock \emph{Int. J. Uncertain. Fuzziness Knowl. Based Syst.}, 10\penalty0
  (5):\penalty0 557--570, 2002.
\newblock \doi{10.1142/S0218488502001648}.
\newblock URL \url{https://doi.org/10.1142/S0218488502001648}.

\bibitem[Wang and Strong(1996)]{DBLP:journals/jmis/WangS96}
Richard~Y. Wang and Diane~M. Strong.
\newblock Beyond accuracy: What data quality means to data consumers.
\newblock \emph{J. Manag. Inf. Syst.}, 12\penalty0 (4):\penalty0 5--33, 1996.
\newblock \doi{10.1080/07421222.1996.11518099}.
\newblock URL \url{https://doi.org/10.1080/07421222.1996.11518099}.

\bibitem[Whang et~al.(2023)Whang, Roh, Song, and
  Lee]{DBLP:journals/vldb/WhangRSL23}
Steven~Euijong Whang, Yuji Roh, Hwanjun Song, and Jae{-}Gil Lee.
\newblock Data collection and quality challenges in deep learning: a
  data-centric {AI} perspective.
\newblock \emph{VLDB Journal}, 32\penalty0 (4):\penalty0 791--813, 2023.
\newblock \doi{10.1007/S00778-022-00775-9}.
\newblock URL \url{https://doi.org/10.1007/s00778-022-00775-9}.

\bibitem[Zha et~al.(2023)Zha, Bhat, Lai, Yang, Jiang, Zhong, and
  Hu]{DBLP:journals/corr/abs-2303-10158}
Daochen Zha, Zaid~Pervaiz Bhat, Kwei{-}Herng Lai, Fan Yang, Zhimeng Jiang,
  Shaochen Zhong, and Xia Hu.
\newblock Data-centric artificial intelligence: {A} survey.
\newblock \emph{CoRR}, abs/2303.10158, 2023.
\newblock \doi{10.48550/ARXIV.2303.10158}.
\newblock URL \url{https://doi.org/10.48550/arXiv.2303.10158}.

\end{thebibliography}
\newpage
\appendix
\section{Definitions and Assessment Challenges of Data Quality Dimensions}
\label{appendix}

We define 29 well-known data quality dimensions. 
The set of dimensions and their definitions are taken from our Data Quality Glossary~\cite{mohammed_glossary_2024}, which was compiled through a thorough literature study. 
This appendix extends those definitions and in particular discusses the challenges of assessing data quality along the individual dimensions. 
These discussions form the basis of scoring the importance of each of the facets, as explained in Section~\ref{sec:facets_application}.

\subsection{Accessibility}
\dq{Accessibility} has technical, organizational, financial, and legal perspectives.
Technical \dq{accessibility} ensures sufficient resources, such as compute power or network bandwidth, at each point of processing to allow smooth and fast access.
Organizational \dq{accessibility} allows users without technical knowledge or with disabilities to access the data easily~\cite[p. 34]{batini_data_2016}.
Legal \dq{accessibility} results from the licensing of legally protected data, which allows for its continued use.
Finally, financial \dq{accessibility} can be achieved through reasonable or waived usage fees.

\paragraph{Assessment challenges.}
From a technical perspective, various test scenarios must be created for the assessment to show how resilient the technologies are under full load or under the influence of disruptions, such as power outages or storage medium failure.
From an organizational point of view, assessment requires designing a user study that includes diverse user groups with different levels of technical understanding and abilities.
From a legal perspective, it requires a law expert who understands the various licensing terms and conditions and verifies their compliance with applicable laws, such as the GDPR\@.
The challenges from a financial perspective include evaluating different usage models and their appropriateness for different user groups, such as individuals, students, and small and large organizations.
\addsingletablerow{-}{+}{++}{-}{++}

\subsection{Accuracy}
\dq{Accuracy} describes the correspondence between a phenomenon in the world and its description as data~\cite{batini_data_2016}.

\paragraph{Assessment challenges.}
\dq{Accuracy} can be assessed at an individual data point, column, or row level, or for the entire considered data.
Therefore, the level of granularity needs to be defined.
If the granularity is not defined, the assessment involves determining the degree of correspondence between the data values and their empirically ascertainable, correct values at each level of granularity.
Thus, the assessment is particularly objective and involves no subjective input.
To assess \dq{accuracy} at higher levels of granularity, such as rows or the overall data, an aggregation of the individual accuracy results is necessary.

A key challenge in this process is obtaining knowledge of the correct or true value(s), which is essential regardless of whether the assessment is binary (equal or unequal) or based on nuanced comparison functions, such as measuring similarity~\cite[p. 20]{batini_data_2006},~\cite{DBLP:journals/csur/BatiniCFM09}. 
Thus, external data sources are needed for the assessment\@.
In addition to leveraging existing reference data, error detection and cleaning methods, such as NADEEF\cite{DBLP:conf/sigmod/DallachiesaEEEIOT13} or HoloClean~\cite{DBLP:journals/corr/RekatsinasCIR17}, can be employed to identify and correct data errors. 
These processes generate a series of data transformations, which should be documented in the metadata, including the applied methods and the corresponding changes made to the data.

The system in which the data are located can also impact \dq{accuracy}.
For instance, system crashes or bugs may alter or corrupt the data. Consequently, an additional challenge is to incorporate information about system robustness, data replication strategies, and recovery processes to ensure data integrity.
\addsingletablerow{++}{+}{+}{-}{-}

\subsection{Added-value}
The \dq{added-value} of data refers to the ability to beneficially utilize data in a use case~\cite{10.1145/505248.506010}.
Data are beneficially utilized if their use results in a profit (monetary, knowledge) for the data owners, or it fulfills a specific task, such as enabling a desired level of model prediction accuracy when used as training data.

\paragraph{Assessment challenges.}
Assessing benefits, especially intangible gains like knowledge, is inherently complex.
Distinguishing the unique contribution of data from other influencing factors in achieving these outcomes poses a significant challenge.
Considering a downstream task, such as an ML context, the assessment of \dq{added-value} can be more straightforward when aligned with measurable outcomes, such as a predefined threshold of prediction accuracy.
\addsingletablerow{-}{-}{-}{++}{++}

\subsection{Appropriate amount of data}
The \dq{amount} of data describes the size of the data that is appropriate to fulfill a specific task~\cite{10.1145/505248.506010, DBLP:conf/iq/NaumannR00}.

It can be too small or too large; for example, a certain amount of training data are needed to adequately train an ML model.
Conversely, an excessive amount of data, such as unnecessarily high-resolution image files, can lead to data management issues.

\paragraph{Assessment challenges.}
Assessing the required amount of data for specific tasks varies significantly based on context.
The classification of what is ``appropriate'' must be defined in advance to reflect the requirements and constraints of the specific application.
This can be assessed with measures such as the size of the data (e.g., measured by bytes or rows) or by an expert user who performs the part of the evaluation.
\addsingletablerow{-}{-}{-}{++}{+}

\subsection{Balance}
\label{app:balance}
The \dq{balance} of data considers the distribution of the contained data points.
Data are balanced if the data points within the represented range of values are equally distributed in relation to each other~\cite{DBLP:journals/tkde/HeG09}.

For example, in a balanced dataset that divides clients into age groups, clients of all ages should be represented in equal numbers.
This does not mean that all age groups of the total population must be included (see \hyperref[sec:dq_diversity]{\dq{diversity}}).

\paragraph{Assessment challenges.}
A challenge in assessing the \dq{balance} of data lies in evaluating data with numerous attributes. 
Assessing the balance of data with multiple attributes may reveal varying degrees of \dq{balance} in different attributes, which must be aggregated appropriately.
Considering an underlying task, the relevance of the imbalance of individual attributes to fulfilling the task must be assessed.
\addsingletablerow{++}{-}{-}{+}{-}

\subsection{Believability}
\dq{Believability} describes the degree to which the available information is regarded as correct~\cite{DBLP:conf/iq/NaumannR00}~\cite[p. 424-426]{batini_data_2016}.

\paragraph{Assessment challenges.}
\dq{Believability} cannot be represented exclusively as a statistical quantity.
Rather, input of users is required, in which they express their opinions about the data or its source.
Also, relevant for the assessment are information about the provenance of the data~\cite[p. 424]{batini_data_2016}, and further documentation about the data.
\addsingletablerow{+}{++}{-}{-}{++}

\subsection{Completeness}
\dq{Completeness} refers to the extent to which data, including entities and attributes, are present according to the data schema~\cite{10.1145/505248.506010}.

\paragraph{Assessment challenges.}
The assessment involves two perspectives.
The first quantifies the extent of missing values inside the data, which is a straightforward task when such values are explicitly identified or represented by conventional placeholders, like ``NaN''.
However, the placeholders for missing values are not always known, they are ``hidden missing values''~\cite{FAHES2018}.
A common example is using specific but arbitrary values to fill missing entries, such as representing a missing date with~\texttt{1900-01-01}. 
Identifying these hidden missing values necessitates prior knowledge of how they are encoded in the data.

The assessment regarding the second perspective includes the quantification of absent tuples that would match the data model schema (open world assumption)~\cite[p. 29]{batini_data_2016}.
To quantify this type of \dq{completeness}, reference data or metadata for the given task is needed~\cite[p. 25]{batini_data_2006}.

Missing tuples can also result from previous transformation strategies, such as deleting them if they contain missing values.
Therefore, transformations must also be part of the metadata and considered during the assessment.
Additionally, the system that stores and processes the data can also cause missing values due to certain failures, such as crashes or bugs.
Thus, similar to the assessment of \dq{accuracy}, the system's recovery process and data replication strategies must also be part of the assessment.
\addsingletablerow{++}{+}{+}{-}{-}

\subsection{Concise Representation}
\dq{Concise representation} considers the form in which data are represented~\cite{DBLP:conf/iq/NaumannR00}.
Concise data are presented suitably and recognizably, depending on the intended use~\cite[p. 45]{batini_data_2006}.

An example is storing timestamps with millisecond precision, such as \texttt{2024-03-01 12:00:00.123}, in data where only minute-level accuracy is needed for a given use case.
Thus, the data are unnecessarily verbose.
Simplifying this to \texttt{2024-03-01 12:00} improves conciseness, making the data more practical and easier to work with for its intended use.

\paragraph{Assessment challenges.}
The assessment of \dq{concise representation} is typically user-specific and context-dependent.
The data representation may be appropriate in one context but inappropriate in another.
The extent to which users find the data concisely represented depends on their individual experience.
\addsingletablerow{+}{-}{-}{++}{++}

\subsection{Consistency}
Data are consistent if all conditions imposed on the state of the data are met.
Consistency conditions can include integrity constraints, such as data types, value ranges, dependencies, or relationships across data sources~\cite[p. 35]{batini_data_2016}.

Examples of a lack of consistency include different date formats in a single column, different cities for the same zip code, or purchase orders with invalid customer numbers.

\paragraph{Assessment challenges.}
A primary challenge in the assessment of the \dq{consistency} is determining and understanding the necessary conditions specific to the data, which can be complex, especially for data with diverse characteristics.
Even if constraints are known (through the metadata), it can be challenging to actually find the corresponding violations in the data~\cite{PenaVio20,KhayyatDansing15}.
\addsingletablerow{++}{-}{-}{-}{-}

\subsection{Consistent Representation}
Data are consistent in their representation if no attribute (column) contains two or more unique values that are semantically equivalent (e.g., New York vs. NYC or \texttt{2024-1-12} vs. \texttt{2024-12-1})~\cite{10.1145/505248.506010, DBLP:journals/datascience/CaiZ15, shah_how_2024}.

\paragraph{Assessment challenges.}
A key assessment challenge is identifying semantic equivalence in various representations, which demands syntactic and semantic analysis.
This matching task becomes more complex with large and diverse data, requiring sophisticated automated methods for handling scale and complexity.
The origin of the data must also be considered, as the semantics of individual values depend on the source from which the data is generated. For instance, values within a single domain may be considered semantically equivalent. 
However, when values originate from different domains, they may no longer share the same semantic equivalence.
\addsingletablerow{++}{++}{-}{-}{-}

\subsection{Cost}
The cost of data includes both the monetary costs incurred in generating or acquiring and permanently storing the data, and the personnel costs incurred in acquiring and preparing the data.
Costs may be calculated for the entire considered data or per query to the data~\cite{DBLP:conf/iq/NaumannR00, DBLP:journals/jmis/WangS96}.
Examples are data annotation costs incurred by data stewards or crowd-workers, purchase of data from data brokers or data markets, and storage in the cloud.

\paragraph{Assessment challenges.}
A challenging aspect of \dq{cost} assessment is considering the variety of cost factors.
The duration of data preparation activities can significantly impact overall costs, especially when considering personnel and opportunity costs.
The context determines to what extent the data should be cleaned and how long it should be stored.
While personnel costs for data cleaning are significant, other elements, such as data acquisition, storage, processing, technology and tools investment, data security and compliance costs also play a crucial role.
The scalability of these costs with increasing data volumes presents another challenge: managing, storing, and processing large datasets can lead to substantial cost escalations.
\addsingletablerow{+}{+}{++}{++}{++}

\subsection{Diversity}\label{sec:dq_diversity}
We adopt the \textit{richness} definition of \textit{diversity}, which is known from ecology and measures the number of different species.
Data are diverse if each entity type of the total set occurs at least once.
The data aim to reflect the \textit{diversity} of entity types from the total set, i.e., containing all relevant variants~\cite{magurran_measuring_2021}.

For example, if an employee database (total set) consists of male and female employees. 
The data are diverse if they contain at least one female employee and one male employee from each department. 
Note that we call data balanced (see Section~\ref{app:balance}) if the same number of male and female employees appear for each department.

\paragraph{Assessment challenges.}
The key challenge is defining and identifying the relevant entity types within the data: for instance, humans can specify for which attributes diversity (gender, age, etc.) is required. 
In addition, when diversity is measured for many attributes and their combinations, determining it can be computationally challenging~\cite{asudeh2019assessing}.

The task is complicated by the need for domain-specific knowledge and the potential vastness or ambiguity of the entity range. Even if one regards only values that are present, checking for all combination is computationally expensive~\cite{asudeh2019assessing}.
Additionally, comparing the data's diversity against a potentially vast, poorly defined, or evolving ``total set'' is a significant analytical challenge.
\addsingletablerow{++}{-}{-}{-}{+}

\subsection{Documentation degree}
Data are well-documented if relevant, complete and correct structured metadata and a textual description are available~\cite{DBLP:conf/iq/NaumannR00, DataCards22}.
Typical metadata includes the volume of the data, its syntactic schema~(data types) and its semantic  schema~(table and column names), statistics, information about its provenance and any transformation that has been performed so far.
Textual descriptions, formalized in so-called data sheets, include the data’s purpose and previous use(s).

\paragraph{Assessment challenges.}
In addition to simply checking the availability of documentation, the assessment should include the evaluation of the metadata on a syntactic and semantic level.
This includes ensuring the \dq{completeness} and \dq{accuracy} of metadata, encompassing various elements from technical schemas to transformation history.
The relevance and quality of textual descriptions, such as data sheets outlining the data's purpose and usage history~\cite{DataCards22}, are equally important, yet subjective.
Challenges are compounded by the lack of standardization in documentation formats, the evolving nature of data, and the necessity for specific technical expertise to accurately assess technical details.
Additionally, understanding the provenance of the data, keeping \dq{documentation} updated with ongoing changes, and addressing legal and ethical considerations add complexity to the task.
\addsingletablerow{++}{++}{-}{-}{++}

\subsection{Ease of manipulation}
Data are easily manipulable if changes or additions can be performed intuitively or without prior knowledge (data in an Excel spreadsheet vs.\ data on a website)~\cite{DBLP:journals/jmis/WangS96, 10.1145/505248.506010}.
\dq{Ease of manipulation} can be viewed from both a positive and a negative perspective:
On the one hand, there is a risk that data will -- intentionally or unintentionally -- be falsified~(negative case).
On the other hand, manipulable data can be easily adapted for legitimate individual purposes (positive case).

\paragraph{Assessment challenges.}
Assessing \dq{ease of manipulation} must balance between ensuring the flexibility for legitimate modifications and safeguarding against unauthorized alterations.
This evaluation process defines ``ease'' in data manipulation, which varies based on the technical format, the users' skill levels, and the required tools.
A central challenge lies in distinguishing positive uses, such as adapting data for valid individual purposes, and negative scenarios, like intentional falsification.
\addsingletablerow{+}{-}{++}{-}{++}

\subsection{Efficiency}
The \dq{efficiency} of data measures the effectiveness with which various processes or algorithms can be executed on the data~\cite{ISO25024}.
Factors affecting response time and thus \dq{efficiency} include network traffic, computational complexity, data storage mechanisms, and the volume of the data itself.
Efficient data are characterized by their ability to be processed with minimal delay.

\paragraph{Assessment challenges.}
Assessment challenges include understanding the nuanced interaction between processes and algorithms on response time to predict the latency in advance.
\addsingletablerow{+}{-}{++}{-}{-}

\subsection{Portability}
According to GDPR, \dq{Portability} is a required property of data and describes the ability to transfer structured data reliably and securely from one system to another.
Portable data are formatted according to common standards, such as JSON, CSV or XML~\cite{kuebler-wachendorff_right_2021}.
A simple transferal of personal data from a social network to an external data storage device would be an example of good \textit{portability}.

\paragraph{Assessment challenges.}
Assessing \dq{portability} requires ensuring that data formats are universally compatible and adhere to common standards.
The existence of different communication protocols between the systems adds to the complexity of the assessment.
This assessment must address security and privacy concerns to maintain data integrity and comply with privacy regulations, particularly under GDPR.
\addsingletablerow{++}{-}{++}{-}{-}

\subsection{Precision}
The \dq{precision} comprises three perspectives.
For one, \dq{precision} reflects the consistency of data recorded repeatedly under unchanged conditions, distinct from data \dq{accuracy}.
For instance, if a hospital patient’s vital signs are measured consistently every 120~seconds, this is an example of high \dq{precision}.
However, such regular measurement intervals do not necessarily reflect the \dq{accuracy} of the data, such as potential measurement errors.

\dq{Precision} also pertains to the level of detail in information.
For instance, a form might request a year of birth or an exact birthdate, where the latter represents higher precision.
This perspective also applies to numerical values, such as recording a temperature as 22°C versus 22.34°C, where the latter provides greater numerical precision due to the inclusion of decimals.

A further dimension of precision is the accuracy in categorizing predefined value classes, such as distinguishing between ``navy blue'' and ``midnight blue'' when specifying colors of products.

\paragraph{Assessment challenges.}
The assessment challenges include defining the required level of detail of measurements.
Similarly, the differentiation of individual categories must be appropriately defined.
\addsingletablerow{++}{-}{-}{-}{-}

\subsection{Privacy}
Data are private if the individuals described in the data have control over and access to that data~\cite{10.1145/505248.506010}.
Private data protects the user's right to informational self-determination.
The legal protection of \dq{privacy} can be ensured from an organizational and technical perspective.

Organizational \dq{privacy} can be established through consent declarations by users, which can prohibit the entire use of the data or contain instructions for use, such as task-related access.

For the technical establishment of \dq{privacy}, the data can, for example, be encrypted or anonymized by privacy preserving techniques~\cite[p. 225]{batini_data_2006},~\cite{hutchison_differential_2006, DBLP:journals/ijufks/Sweene02}.

\paragraph{Assessment challenges.}
Ensuring that individuals described in the data maintain control and access over their data while balancing this with the legitimate use of the data is a key challenge in assessing privacy.
Only humans can specify, which attributes should be kept private.
Organizational privacy, on the other hand, relies heavily on interpreting and implementing user consent declarations effectively, a task that requires precision to respect user choices without limiting data accessibility.
Technically, implementing privacy measures like encryption or anonymization poses challenges in selecting suitable techniques and assessing their impact on for data accessibility, the accuracy, integrity constrains or the needed storage.
Additionally, keeping up with evolving legal standards for data privacy and adapting to global variations in privacy laws adds complexity to compliance efforts.
\addsingletablerow{-}{++}{+}{-}{+}

\subsection{Recoverability}
Data show \dq{recoverability} if, despite system errors or data carrier loss, they can be re-created and the previous data quality can be guaranteed~\cite{DBLP:conf/qsic/MoragaMCC09}.

\paragraph{Assessment challenges.}
The assessment of \dq{recoverability} involves evaluating the effectiveness and robustness of data backup and recovery processes, ensuring the existence of backups and their ability to restore data quickly and accurately.
Testing for various failure scenarios is challenging due to the unpredictability and diversity of potential issues.
Another critical aspect is ensuring the integrity and quality of data post-recovery, verifying that the restored data matches the original.
\addsingletablerow{-}{-}{++}{-}{-}

\subsection{Relevancy}
\dq{Relevancy} describes the extent to which data are applicable and helpful for a given task~\cite{DBLP:journals/jmis/WangS96}.
For example, in an online store, the name and price of an article are relevant for the comparability of products.
On the other hand, the number of people involved in manufacturing the individual products can be of little relevance, depending on the use case.

In the context of machine learning, during the training phase of a model, the \dq{relevancy} of attributes in the training data vary in their impact on achieving high prediction accuracy on the test data.

\paragraph{Assessment challenges.}
Determining what constitutes relevant data varies considerably based on user perspectives and application-specific needs.
The dynamic nature of \dq{relevancy}, evolving with changing user requirements, market trends, and legal standards, adds to the complexity of maintaining up-to-date relevance assessments.
The assessment also involves balancing the need for complete information against the risk of including unnecessary data, which can complicate data management and violate legal requirements.
\addsingletablerow{+}{-}{-}{++}{++}

\subsection{Reliability}
\dq{Reliability} describes the extent to which the data can be trusted: the information represented is correct~\cite[p. 38]{batini_data_2006}\cite{DBLP:conf/iq/NaumannR00}.

\paragraph{Assessment challenges.}
Key challenges are verifying the credibility of data sources, maintaining data integrity and consistency.
Challenges include ensuring data completeness, managing biases, and staying current with the data's temporal relevance.
Additionally, a domain expert can also be part of the assessment, who assesses the data from a semantic and syntactic perspective, which is associated with a subjective factor.
\addsingletablerow{++}{++}{-}{-}{++}

\subsection{Representativity}
\dq{Representativity} aims to ensure that the (statistical) characteristics of the reference data are present in the considered data~\cite{10bd9b32-8b48-36d1-a9cc-51fbb4bf7d53, DBLP:journals/corr/abs-2203-04706}.

For example, let a dataset (total set) consist of 70~male and 30~female students. Of the male students, 40 study art and 30 study history, while 15 of the female students study art and the remaining 15 study history.
Based on the previously stated total, a sample of the dataset would be representative if it consisted of 9 female art and history students, 24 male art students and 18 male history students.
This dataset is statistically representative because the relative ratios between students of the same sex and between students of the same major are identical when compared to the total set.  

\paragraph{Assessment challenges.}
Challenging to the assessment of \dq{representativity} is comparing the distribution of key characteristics in the given data against those of the total population.
This requires first defining the total population. 
For the comparison, complete raw data is not necessary -- summary statistics or data distributions are often sufficient.
However, even in aggregated formats, comparing the given data with reference data necessitates the use of data matching strategies. 
To enable this, the data must be in the appropriate format.
\addsingletablerow{++}{-}{-}{-}{-}

\subsection{Reputation}
\label{app:reputation}
The \dq{reputation} of data describes the trustworthiness of the data source and the content~\cite{10.1145/505248.506010}.
Data and data sources have a high \dq{reputation} if they have already proven high quality in the past and over some time.
Conversely, if no or only poor experience has been gained with them in the past, their \dq{reputation} is low.
In particular, if other data quality dimensions, such as \dq{accuracy}, cannot be adequately measured, \dq{reputation} can also be understood as the expected quality of data.

\paragraph{Assessment challenges.}
Key assessment challenges are the subjective nature of \dq{reputation}, the necessity of analyzing historical data for quality trends~(see Section~\ref{app:traceability}), and the dynamic nature of reputation over time.
External factors such as media influence, public opinion, and cultural differences can significantly sway the perceived \dq{reputation}, adding to the complexity.
Moreover, the risk of confirmation bias and the need to balance historical \dq{reputation} with current performance make the assessment even more complex.
\addsingletablerow{-}{++}{-}{-}{++}

\subsection{Security}
Data \dq{security} describes the extent of protection against unauthorized access to data~\cite{10.1145/505248.506010}.
Systems must guarantee correct access management; to maintain this guarantee, a system's functional security is also relevant so that in the event of a functional failure, the system will still enter a defined state in which the data \dq{security} is guaranteed.

For example, a customer of an online store should have access only to their orders and not to other orders of that store.

\paragraph{Assessment challenges.}
An essential assessment challenge is ensuring that systems maintain \dq{security} during diverse functional failures.
To test this, automated and user-guided tests are needed.
Assessing compliance with diverse legal and regulatory standards further complicates the process, as does addressing the human factor, which introduces significant vulnerabilities.
Resource constraints add to the complexity, especially in implementing continuous monitoring and rapid response systems.
\addsingletablerow{-}{-}{++}{-}{+}

\subsection{Timeliness}
\dq{Timeliness} describes the difference in time between an electronically captured event in the real world and its digital representation in the data, considering the task at hand~\cite[p. 38]{batini_data_2006}.
Changes can result from new data being captured~(e.g., a sale), existing data becoming outdated due to real-life events~(e.g., a customer moving), or data being deleted (e.g., a company going bankrupt). 

\paragraph{Assessment challenges.}
The key assessment challenge is defining an acceptable time frame for various tasks and classifying how long data are considered up-to-date.
This can vary depending on the application.
\addsingletablerow{+}{++}{-}{++}{-}

\subsection{Traceability}
\label{app:traceability}
\dq{Traceability} describes the ability to trace the provenance of data, including their origin and all transformations performed on them~\cite{DBLP:journals/jmis/WangS96, DBLP:journals/vldb/HerschelDL17}.

Tracing data facilitates the restoration of data to a previous state, e.g., using common version control systems or appropriate data sheets for documentation so that either the current version is replaced or different versions of the data exist in parallel.
High \dq{traceability} is also useful for assessing other data quality dimensions, such as \dq{reputation} (see Section~\ref{app:reputation}).

\paragraph{Assessment challenges.}
Information on the provenance of the data must be available and correct.
Appropriate software or established processes are required to make this information accessible in the long term.
Users from different user groups should be able to access the required information easily.
\addsingletablerow{+}{++}{+}{-}{-}

\subsection{Transparency}
The \dq{transparency} dimension measures the extent to which stakeholders can access all data-related information, including the origin of the data, data collection strategies, and the transformations applied to them~\cite{DBLP:journals/jdiq/BertinoKS19, DataCards22}.

\paragraph{Assessment challenges.}
Assessment challenges include determining whether the disclosed information has been prepared in an understandable way for various stakeholders.
Another point is the assessment of the compliance of the disclosure.
It may be that the associated information may not be disclosed or only partially disclosed.
\addsingletablerow{+}{++}{-}{-}{++}

\subsection{Understandability}
The \dq{understandability} describes the extent to which a user can semantically comprehend the information represented by the data~\cite{DBLP:conf/iq/NaumannR00}.

For example, an online store’s data are understandable if the full names of articles are listed, so customers can immediately recognize them.
Understandability is impaired, for instance, if only an article number is listed instead of the full name.

\paragraph{Assessment challenges.}
Assessing the \dq{understandability} of data encompasses numerous challenges, primarily stemming from the diversity of user backgrounds and the inherent complexity of information.
Tailoring data presentations to be comprehensible across varied educational, cultural, and professional spectra while maintaining \dq{accuracy} and avoiding oversimplification is a key challenge.
Using technical language and selecting appropriate visualization techniques requires careful consideration to avoid misinterpretation and cognitive overload.
Additionally, ensuring cultural and contextual sensitivity in data presentation is essential, especially for global audiences.
\addsingletablerow{++}{-}{-}{-}{++}

\subsection{Uniqueness}
\dq{Uniqueness} measure whether each entity in the real world is represented at most by one entry in the data, meaning there are no duplicates~\cite{DBLP:journals/csur/BatiniCFM09}.

For example, the same customer shall appear only once in a customer database.

\paragraph{Assessment challenges.}
Measuring \dq{uniqueness} raises two main challenges.
The first is a succinct definition to determine when two entries are considered duplicates.
The principle of exact duplicates exists in the literature, where entries must be completely identical.
On the other hand, entries can be classified as duplicates based on a similarity function, even if they do not match exactly (fuzzy duplicates)~\cite{naumann_introduction_2010,Christen12}.
The granularity must also be defined, i.e., whether the \dq{uniqueness} is measured at value, row, or column level or across entire datasets.

Second, it can be computationally expensive to discover all duplicates in data and thus determine their \dq{uniqueness}, so estimations can help~\cite{Heise14}.
\addsingletablerow{++}{-}{-}{-}{-}

\end{document}